\documentclass[conference]{IEEEtran}
\IEEEoverridecommandlockouts

\usepackage{cite}
\usepackage{graphicx}
\usepackage{url}
\usepackage{amssymb}
\usepackage{amsmath}
\usepackage{gensymb}
\usepackage{siunitx}
\usepackage{booktabs}
\usepackage{tabularx}
\usepackage[hidelinks,bookmarks=false,pdfpagelabels=false]{hyperref}
\usepackage{orcidlink}

\begin{document}

\title{Sustainable Edge Vision via Empirically Calibrated DVFS: Eliminating Thermal Throttling on Passively Cooled Hardware}

\author{
    \IEEEauthorblockN{Aayush Marasini\,\orcidlink{0009-0006-2143-2306} and Zhaoxian Zhou\,\orcidlink{0000-0001-6789-166X}}
    \IEEEauthorblockA{\textit{School of Computing Sciences and Computer Engineering} \\
    \textit{University of Southern Mississippi}, Hattiesburg, MS, USA \\
    Email: \{aayush.marasini, zhaoxian.zhou\}@usm.edu}
}

\maketitle

\begin{abstract}
Passive cooling eliminates the energy overhead and mechanical failure modes of fans, making it attractive for edge deployment, yet sustained Deep Neural Network (DNN) inference on passively cooled edge Systems-on-Chip (SoCs) is bottlenecked by thermal throttling. To address this, we propose an empirically calibrated, state-aware Dynamic Voltage and Frequency Scaling (DVFS) scheduler. Unlike heuristic-driven controllers, our methodology utilizes time-domain guards and absolute temperature bounds, with derivative triggers acting as safeguards against sharp thermal spikes. Evaluated on a passively cooled Raspberry Pi 5 running YOLOv8n, our scheduler eliminates all observed thermal throttling events during sustained 30-minute workloads. It outperforms a temperature-only reactive baseline by achieving a 6.8\% higher frame rate (Cohen's d = 8.73) while consuming 1.9\% less energy per frame. Furthermore, our optimized passive scheduling surpasses an actively cooled reference system in energy efficiency (Joules/frame), though active cooling remains superior for raw throughput. Through isolated ablations, we show that the dwell guard is necessary for run-to-run reproducibility. Finally, exploratory boundary probes indicate that the passive operating envelope closes at ambient temperatures $(\ge \qty{27}{\celsius})$ where nonlinear leakage defeats DVFS-based control. These results indicate that, within the mapped envelope, correct scheduling can make mechanical cooling unnecessary for sustained edge inference on this platform.

\end{abstract}


\section{Introduction}
\label{sec1}
The energy footprint of always-on edge inference is a growing sustainability concern, and cooling is a part of that footprint: fans draw continuous power, accumulate dust, and fail mechanically. Passive cooling removes these costs and makes edge deployment cheap and reliable, yet it is exactly what renders sustained inference thermally infeasible. Frameworks such as YOLO have been widely optimized for robotics and embedded systems to enable continuous on-device inference \cite{nizeniecYOLOObjectDetectors2026, jainConvolutionalNeuralNetworks2021}. However, growing model complexity strains the limited compute, memory, and power budgets of edge nodes, and sustained deep-learning workloads rapidly exhaust them \cite{ngoEdgeIntelligenceReview2025, shuvoEfficientAccelerationDeep2022}.\\  
Reactive triggers act too late on high-inertia passive boards, and mobile DVFS targets battery life, not thermal survival. Quantization can also backfire on commodity runtimes. Predictive control is computationally heavy for an SBC, leaving a gap for lightweight, calibrated regulation on such boards (Section~\ref{sec2}). On our platform, sustained YOLOv8n on a fan-less Raspberry Pi 5 throttles 1823 times in 30 minutes at full speed.\\
In this work we build an empirically calibrated, state-aware DVFS scheduler whose thresholds are derived from measured sensor noise, not tuning. While the derivative logic is incorporated as a safeguard for thermal spikes, state transitions during our nominal runs are governed strictly by absolute temperature bounds and dwell constraints. Our work makes the following primary contributions: 
\begin{enumerate}
\item We introduce an empirically calibrated DVFS scheduling methodology for passively cooled edge SoCs, with all thresholds derived from on-device sensor noise and thermal step-response measurements rather than manual tuning, and evaluate it on a Raspberry Pi 5.
\item We establish a rigorous, highly replicable testing framework anchored by frozen SHA256 artifacts, bootstrap statistics, and energy cross-validation.
\item We provide two controlled comparisons that isolate the sources of the gains: a dwell-guard ablation, and a comparison against a temperature-threshold-only reactive baseline that quantifies the value of the time-domain guard stack.
\item We show that, on the evaluated platform, calibrated passive scheduling outperforms an actively cooled reference on energy-per-frame (0.531 vs 0.563 J/frame) by trading peak raw performance for superior energy efficiency, and we provide an initial characterization of the passive operating envelope via exploratory boundary probes, locating the ambient boundary ($\ge\qty{27}{\celsius}$) beyond which passive control fails. 
\end{enumerate}

\section{Related Work}
\label{sec2}
Current literature heavily explores algorithmic compression techniques, predominantly low-bit quantization. Transforming weights and activations from FP32 to INT8 or INT4 is widely cited as a direct mechanism to reduce memory bandwidth and accelerate inference on edge CPUs \cite{ashfaqAcceleratingDeepLearning2022, mahmudovQuantEdgeHybridQuantization2025}. Reported INT8 gains, however, are highly runtime and hardware-specific: Ahn \textit{et al.}~\cite{ahnPerformanceCharacterizationUsing2023} note that while OpenVINO excels on Intel CPUs, they obtain their best speedups on ARM-based Raspberry Pi hardware using TensorFlow Lite. Recent work further exposes a widening gap between simulation and practical deployment for sub-INT8 schemes such as INT4 \cite{koseBridgingGapAI2025}. However, as our research reveals, this hardware-software deployment gap is not limited to sub-INT8. In practice, low-bit precision is bound by the maturity of the software stack and specific hardware extensions (such as ARM's I8MM). Our findings build upon this critical nuance: we observe on our stack that INT8 deployment on certain combinations of runtime and hardware can actually invert the expected benefits, bottlenecking execution units and drastically reducing overall energy efficiency.\\
Dynamic Voltage and Frequency Scaling (DVFS) serves as a critical mechanism for thermal control. Extensive research has been dedicated to optimizing embedded systems for thermal and energy efficiency utilizing DVFS, deep reinforcement learning, adaptive task scheduling, and predictive thermal models \cite{ahmadiEdgeEngineThermalAwareOptimization2024, yatskivPREDICTIVETHERMALMANAGEMENT2026, liEnergyEfficientComputationDVFS2025, jeonPhoenixThermalAwareOnDevice2026}. Other strategies integrate DVFS with early-exiting networks or cloud offloading to reduce the computational burden before critical temperatures are reached \cite{zhangE4EnergyEfficientDNN2025, zhangSparseDVFSSparseAwareDVFS2026}. However, the vast majority of these solutions operate under the assumption of active cooling or rely on instantaneous reactive thresholding \cite{zhangThermalAwareOnDeviceInference2022}. They frequently trigger immediate frequency scaling based on isolated temperature readings or rely on heavy predictive loops that consume the very compute overhead they attempt to save. Closest to our setting, run-time throttle prevention via reinforcement-learning prediction has been demonstrated on actively cooled Jetson boards \cite{nishaRunTimePreventionThermal2024}, and thermo-aware anytime inference on the Raspberry Pi 5 controls the \emph{accuracy} axis, degrading model fidelity under thermal pressure \cite{jacobThermoAwareAnytimeInference}. In contrast, we hold accuracy fixed and regulate only the frequency axis, with all controller parameters calibrated from measured thermal dynamics rather than learned or hand-selected.
Despite these advancements, the current optimization frameworks fail to account for the physical thermal inertia inherent to completely passive systems. In devices relying solely on a passive heatsink, temperature responses lag significantly behind power dissipation. Consequently, reactive temperature triggers are vulnerable to premature escalation and non-reproducible state trajectories on systems with high thermal inertia, an effect we directly observe when our own dwell guard is removed (Section~\ref{sec5}). This destroys run-to-run reproducibility and degrades energy efficiency. By integrating empirically calibrated dwell times ($2\tau_{thermal}$) and sample confirmation counts $(N_{confirm})$, our methodology safely manages delayed thermal trajectories and consistently produced stable state transitions in our evaluation without relying on active cooling or computationally expensive predictive modeling.  
\section{System Design and Configuration Space}
\label{sec3}
\subsection{The Thermal Problem}
\begin{figure}[htbp]
    \centering
    \includegraphics[width=\linewidth]{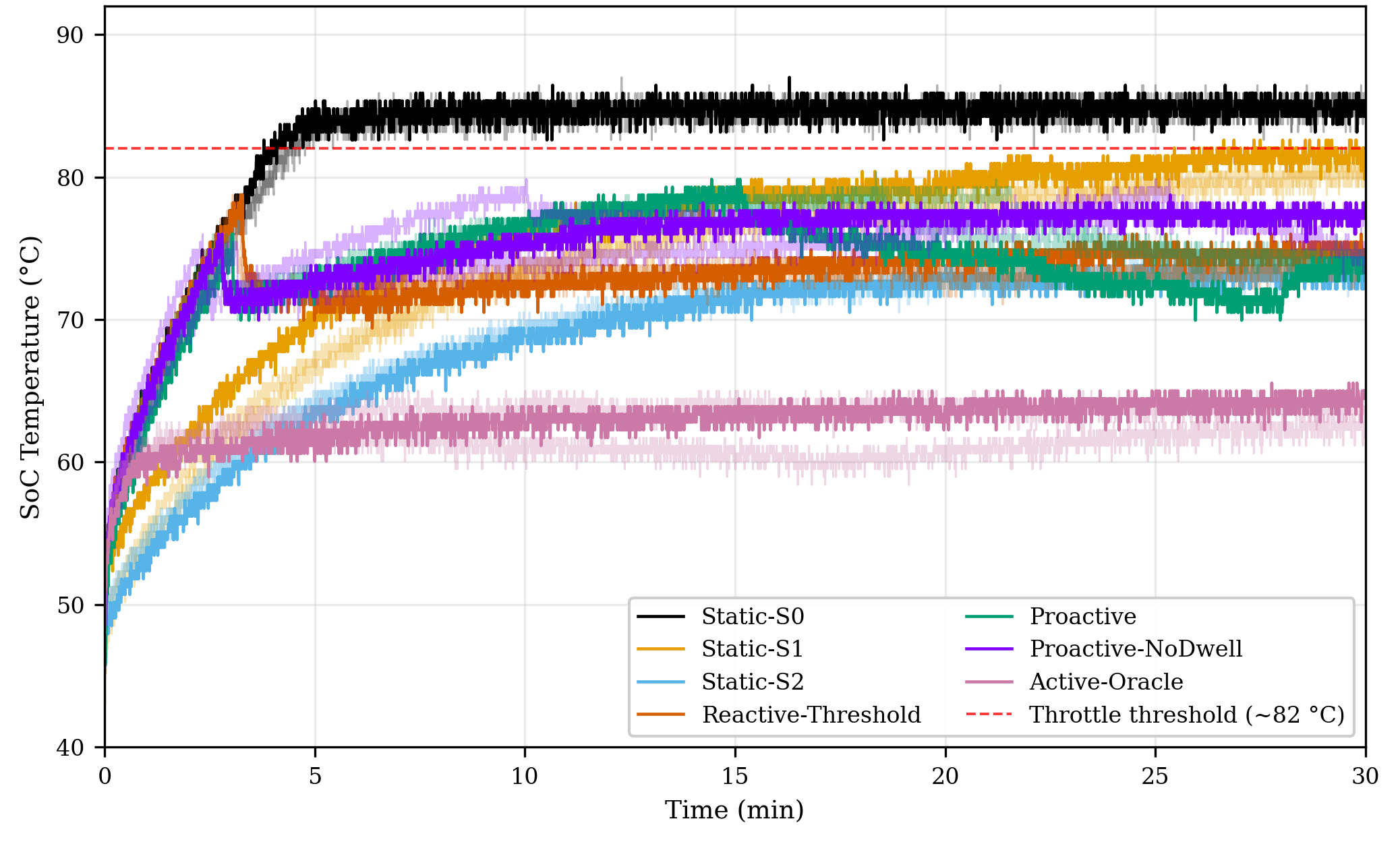}
    \caption{SoC temperature trajectories, all seven configurations, 30-minute sustained inference. Dashed line: $\qty{82}{\celsius}$ margin reference (Pi 5 hardware throttle: $\qty{85}{\celsius}$).} 
    \label{fig:thermal_trajectories}
    
    \end{figure}

Sustained inference on the passively cooled Pi 5 drives the System on a Chip (SoC) into hardware throttling. Sustained inference at full speed (S0) saturates at $\qty{84.8}{\celsius}$ and throttles 1823 times (Table~\ref{tab:main_results}) in a 30-minute run. This plateau sits at the Pi 5's $\qty{85}{\celsius}$ throttle limit, which we observe in the S0 telemetry; transient excursions above it produce the 1823 logged events. Thermal margins in this paper are reported against a conservative $\qty{82}{\celsius}$ reference. As seen in Fig.~\ref{fig:thermal_trajectories}, this is the thermal wall the rest of the paper addresses.

\subsection{DVFS Configuration Space}
DVFS via \texttt{scaling\_max\_freq} on the ondemand governor \cite{pallipadiOndemandGovernor2006} gives us three states: S0 2400, S1 1800, S2 1500 MHz. We ran 5-minute $n{=}3$ profiling runs. Throughput falls \emph{sublinearly} with the frequency cap: a 25\% reduction (S0$\rightarrow$S1) costs only $\approx$15\%
FPS, while board power falls $\approx$26\%. Because power falls faster than throughput, S1 and S2 lower J/frame relative to S0, yielding genuine efficiency gains. Model outputs remain bitwise-identical across states. 

\subsection{Static Thermal Validation}
We ran each state three times for 30 minutes ($n{=}3$). S0 plateaued at $\qty{84.8}{\celsius}$ and throttled, but S1 and S2 did not throttle. These per-state plateaus from Table~\ref{tab:thermal_val} are what empirically set the escalation/recovery thresholds listed in Table~\ref{tab:scheduler_parameters}. 
\section{Methodology}
\label{sec4}
\subsection{Hardware Platform}
All experiments were conducted on a Raspberry Pi 5 Model B (Broadcom BCM2712, Quad-core ARM Cortex-A76 at 2.4~GHz, 8 GB LPDDR4X-4267) running Raspberry Pi OS Lite (Debian GNU/Linux 13 (trixie) aarch64, kernel 6.12.75+rpt-rpi-2712).The active-cooling reference used the official Raspberry Pi 5 Active Cooler (fan + heat-sink). For passive runs the fan was disconnected at its header, leaving only the heatsink. SanDisk SR64G was used for storage. All experiments used the 27 W official Raspberry Pi USB-C power supply connected inline with ChargerLAB POWER-Z KM003C with 1~kS/s sampling. All runs shared one workbench at $23 \pm \qty{2}{\celsius}$ ambient, logged at start and end via a DHT11 sensor \cite{aosong_dht11}. Runs were executed with a $\ge$15-minute cooldown between runs; each run began only after the SoC returned to its idle-temperature band. 

\subsection{Workload and Model}
We used the Ultralytics YOLOv8n \cite{ultralytics_yolov8} model trained on the USA subset of the RDD2022 Dataset \cite{aryaRDD2022MultinationalImage2022}. Data was split 70/10/20 train/val/test $(n = 3363/481/961)$, seed 42, and trained on a Kaggle Tesla T4 (batch 16, 300 epochs). Training stopped early at epoch 218 with the best epoch being 168 with $mAP50=0.533$. The model was deployed via an OpenVINO 2026.0.0 FP32 intermediate representation (IR); deployed mAP50 on the Raspberry Pi was 0.538, a +0.005 difference within run-to-run noise. The test split $n=961$ was stitched into video at 30 FPS (seed 42) via OpenCV. These disjoint images were never seen during training/validation. This 32-second video was looped $\approx$56 times per 30-minute run; the disjoint frames force feature extraction on every frame. 
\subsection{Telemetry}
The telemetry was sampled at \qty{2}{\hertz}, $20\times$ the thermal
characteristic frequency ($1/\tau_{thermal} \approx \qty{0.1}{\hertz}$
for $\tau_{thermal} \approx \qty{10}{\second}$). A single sampling process feeds a queue which has two consumers (CSV writer and scheduler), therefore logging never blocks inference. The SoC temperature, CPU frequency, CPU utilization and the memory utilization percentages are sampled at 2~Hz whereas core voltage and the throttle bits are decimated 5$\times$, yielding a 0.4~Hz effective rate. The throttle status comes from the kernel's \texttt{vcgencmd get\_throttled}. We log the full raw bitmask along with two derived flags: \texttt{throttled\_now} (bit 2, an active thermal throttle event), and \texttt{undervolt\_now} (bit 0, an under-voltage event). Each signal is smoothed with an Exponential Moving Average (EMA), and derivatives ($\dot{T}$, etc.) are computed by a stride-k backward finite difference on the smoothed signal, with $k = 4$ at 2~Hz. The telemetry pipeline creates \texttt{telemetry\_raw.csv}, \texttt{telemetry\_derived.csv}, \texttt{inference\_log.csv}, \texttt{scheduler\_decisions.csv} (for scheduler runs) 
and \texttt{run\_metadata.json}, at a measured overhead of 1.90\% relative FPS reduction (1.33~ms/frame absolute) from paired $n{=}3$, 5-minute benchmarks (Table~\ref{tab:telemetry_overhead}).   

\begin{table}[htbp]
    \centering
    \caption{Telemetry overhead}
    \label{tab:telemetry_overhead}
    \begin{tabularx}{\columnwidth}{X c c}
        \toprule
        Configuration & FPS & std \\
        \midrule
        Inference-only & 14.579 & 0.019 \\
     Inference + telemetry (2~Hz) & 14.302 & 0.067 \\
        \bottomrule
    \end{tabularx}
\end{table}
\subsection{Energy Measurement}
Energy was measured using the inline ChargerLAB POWER-Z KM003C. We applied an active-power filter to retain all the samples where instantaneous board power $P(t) = V_{BUS} \times I_{BUS} > \qty{4.0}{\watt}$ and mean was computed from all the retained samples. The 4 W threshold was selected because the Pi 5 idles at $\approx\qty{2.5}{\watt}$ while the lowest sustained
inference power we observed was $\approx\qty{5.3}{\watt}$ (Static-S2); \qty{4}{\watt}
is a clean separator between the two regimes. The filter compensates for $\pm$30--60~s of manual drift in Power-Z start/stop timing. Throughput was computed from \texttt{inference\_log.csv} as $\mathrm{FPS} = N_{\mathrm{frames}}(t \ge \qty{10}{\second}) / (t_{\mathrm{last}} - t_{\mathrm{first}})$, and energy per frame as $\mathrm{J/frame} = \bar{P}/\mathrm{FPS}$.
Predicted scheduler power (from per-state baselines and time-at-state fractions) matched observations within 1.5\% (proactive) and 3.2\% (reactive).
\subsection{Calibration}
The absolute temperature sensor noise ($\sigma_T$) was measured to be \qty{0.5835}{\celsius} using an idle calibration run and was used for the hysteresis band floor ($\ge3\sigma_T = \qty{1.75}{\celsius}$). The post-EMA derivative noise ($\sigma_{\dot{T}}$) was \qty{0.0759}{\celsius}/s calculated from the calibration run with the derivative computed at EMA stride = 4 and it was used for the proactive trigger threshold ($6.6 \times \sigma_{\dot{T}} = \qty{0.5}{\celsius}/s$). Finally, the heatsink RC time constant ($\tau_{thermal}$) was found to be 10 s. It was found using empirical step-response, from the stress-ng start to the 90\% rise of the $\dot{T}$ signal. It was then used for the dwell time floor ($2 \times \tau_{thermal} = \qty{20}{\second}$).  
\subsection{Scheduler Decision Policy}

The scheduler selects among the three DVFS states defined in Section~\ref{sec3}. The scheduler decision frequency is equal to the telemetry sampling rate (2~Hz). The scheduler parameters are as follows. 
\begin{table}[htbp]
    \centering
    \caption{Scheduler decision parameters. Derivations reference the calibration-phase profiling plateaus, which differ slightly from the final 30-min validation plateaus in Table~\ref{tab:thermal_val}.}
    \label{tab:scheduler_parameters}
    \begin{tabularx}{\columnwidth}{l c >{\raggedright\arraybackslash}X} 
        \toprule
        Parameter & Value & Derivation \\
        \midrule
        $T_{\text{esc}}(S0 \rightarrow S1)$ & 75.0~$^\circ$C & S0 plateau 84.9 $-$ 9.9~$^\circ$C headroom \\ 
        $T_{\text{esc}}(S1 \rightarrow S2)$ & 79.0~$^\circ$C & S1 plateau 81.3 $-$ 2.3~$^\circ$C headroom \\ 
        $T_{\text{rec}}(S2 \rightarrow S1)$ & 71.0~$^\circ$C & 8~$^\circ$C below $T_{\text{esc}}(S1 \rightarrow S2)$ \\ 
        $T_{\text{rec}}(S1 \rightarrow S0)$ & 68.0~$^\circ$C & 7~$^\circ$C below $T_{\text{esc}}(S0 \rightarrow S1)$ \\ 
        $\dot{T}$ trigger & 0.5~$^\circ$C/s & $6.6 \times \sigma_{\dot{T}}$ \\ 
        $T_{\text{floor}}$ (proactive arm) & 65~$^\circ$C & below this, $\dot{T}$ cannot reach threshold within $\tau$ \\ 
        N\_confirm & 3 samples & 1.5~s at 2~Hz \\ 
        Dwell time & 20.0~s & $2 \times \tau_{\text{thermal}}$ \\ 
        \bottomrule
    \end{tabularx}
\end{table}

The scheduler evaluates four conditions in a priority order: 

\textbf{Recovery}: If the current temperature is lower than the recovery temperature ($T<T_{rec}$) for the confirmed samples ($N\_confirm$) AND dwell has elapsed, the scheduler steps down into the lower state.

\textbf{Proactive escalation}: If the current temperature is greater than or equal to the floor temperature ($T\ge T_{floor}$) AND current temperature derivative is greater than threshold derivative ($\dot{T} > \dot{T}_{\text{thresh}}$) for the confirmed samples ($N\_confirm$) AND dwell has elapsed, it steps up the state.

\textbf{Reactive escalation}: If the current temperature is higher than the escalating temperature ($T>T_{esc})$ for the confirmed samples ($N\_confirm$) AND dwell has elapsed, it steps up the state.

\textbf{Hold}: The scheduler holds the current state otherwise. 

We have enforced the following safety invariants.
\begin{itemize}
    \item $\Delta T_{hyst} \ge 3\sigma_T$ for all threshold pairs
    \item $Dwell \ge 2\tau_{thermal}$
    \item N\_confirm = 3 samples required before any transition
    \item No state skipping 
    \item DVFS restored to S0 on clean exit
\end{itemize}
Set-points were fixed once from the calibration-phase plateaus, subject to the floors above, and were not retuned on the evaluation runs.
\subsection{Statistical Methodology}
We replicated each run three times per main condition. We computed percentile-bootstrap confidence intervals (10{,}000 resamples, seed 42) at the 95\% level. Each replication is a 30-minute run plus a cooldown period, bounding total bench time; we compensate for the small $n$ with a paired design, bootstrap intervals, and effect-size reporting. We also calculated paired Cohen's d matched by replication index. We report effect sizes rather than p-values, as the paired effect sizes are large and the sample is small.
\subsection{Reproducibility}
All model weights, dataset partitions, and the benchmark video are locked with SHA256 hashes in \texttt{00\_frozen\_artifacts/SHA256SUMS.txt}. Random seed 42 was used across split, training and calibration. Other conditions were YOLOv8 deterministic = True, CUDNN deterministic = True, CUDNN benchmark = False during training. Per-run metadata (hardware/software versions, ambient temperature) are written to \texttt{run\_metadata.json} in the repository. All code and data are available at \url{https://github.com/Aayush-Marasini/sustained-edge-vision}.
\section{Results}
\label{sec5}
\subsection{Static Baselines}
\begin{figure*}[htbp]
    \centering
    \includegraphics[width=0.75\linewidth]{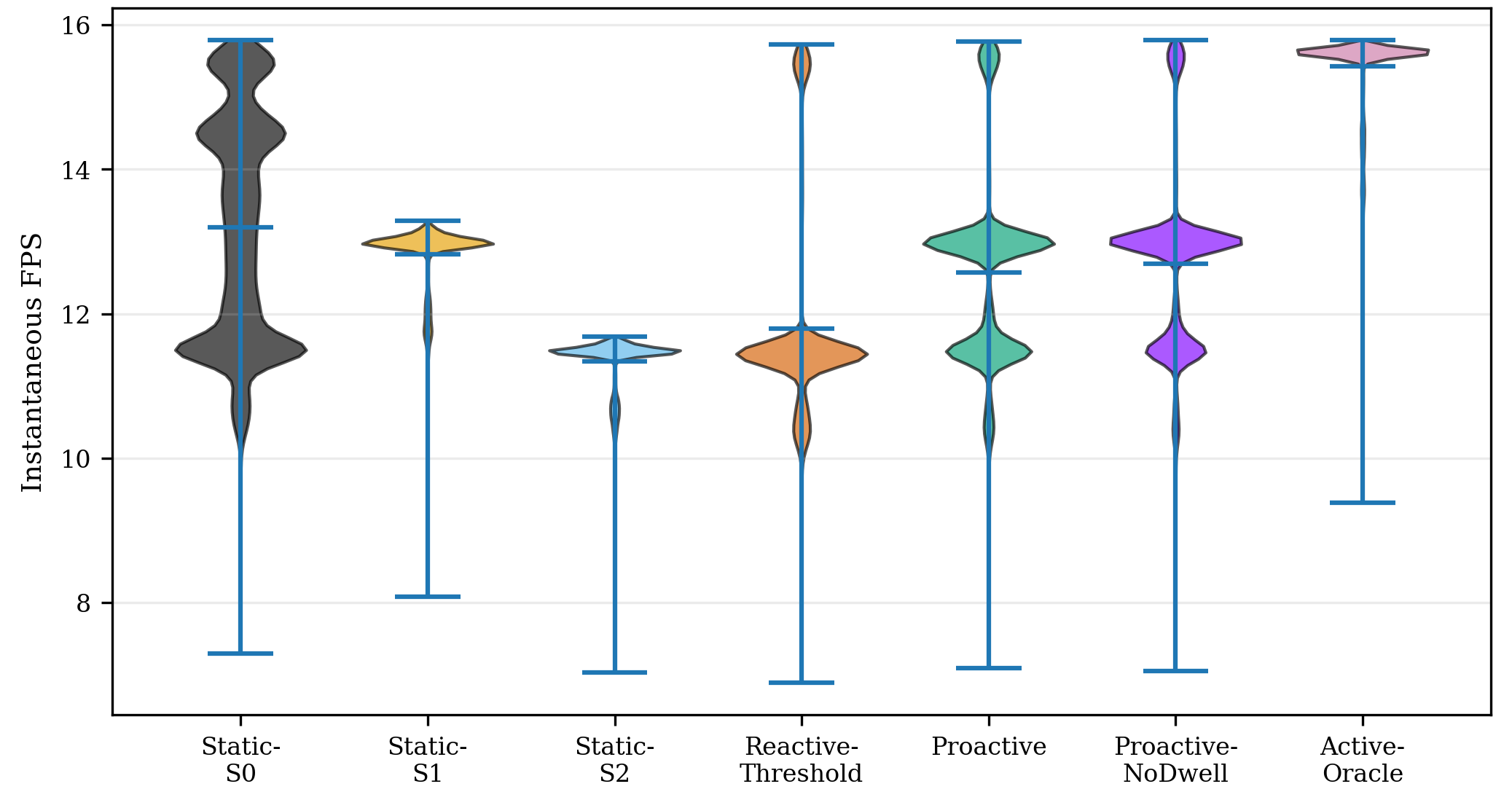}
    \caption{FPS Distribution (all reps pooled, $t > 10$ s)}\label{fig:fps_distribution}
    \end{figure*}

The three static states span the throughput-thermal trade: S0 runs fastest (12.31 FPS) but saturates at $\qty{84.8}{\celsius}$ and throttles 1823 times. S2 stays the coolest at $\qty{73.4}{\celsius}$ at the lowest throughput (10.76 FPS) as seen in Table~\ref{tab:thermal_val}. 
Static S1 sits at $\qty{80.3}{\celsius}$ which is only $\qty{1.7}{\celsius}$ below the $\qty{82}{\celsius}$ reference. So, it avoids throttling at this ambient but with minimal margin. 
Pooled per-frame FPS distributions appear in Fig.~\ref{fig:fps_distribution}.

\subsection{Proactive vs Reactive}
\begin{figure}[htbp]
    \centering
    \includegraphics[width=\linewidth]{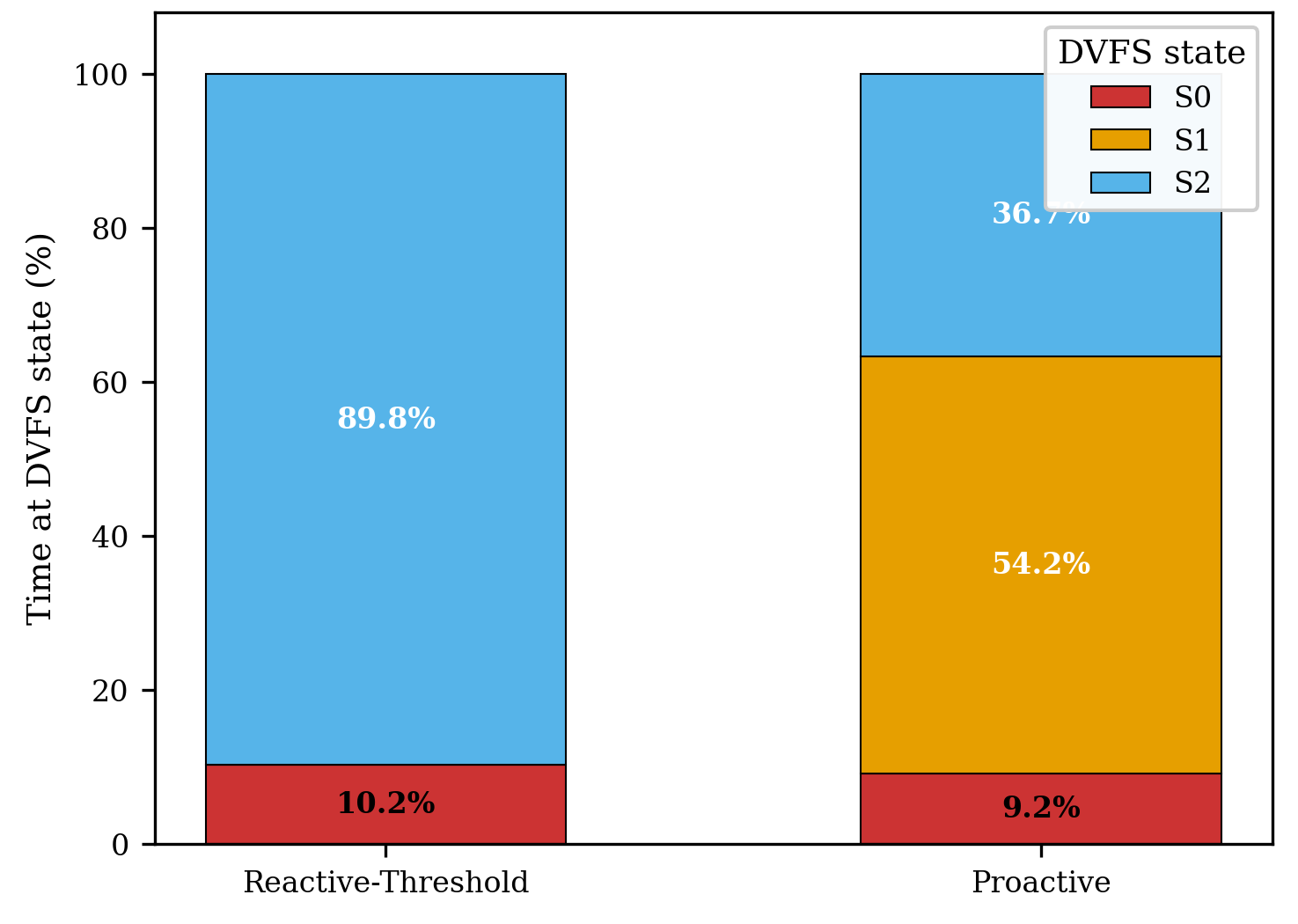}
    \caption{Time-at-state distribution (Reactive vs Proactive). Reactive
    is locked to S0/S2, while Proactive holds the efficient S1 state for
    54.2\% of the run.}\label{fig:time_at_state}
\end{figure}

\begin{figure}[ht]

    \centering
    \includegraphics[width=\linewidth]{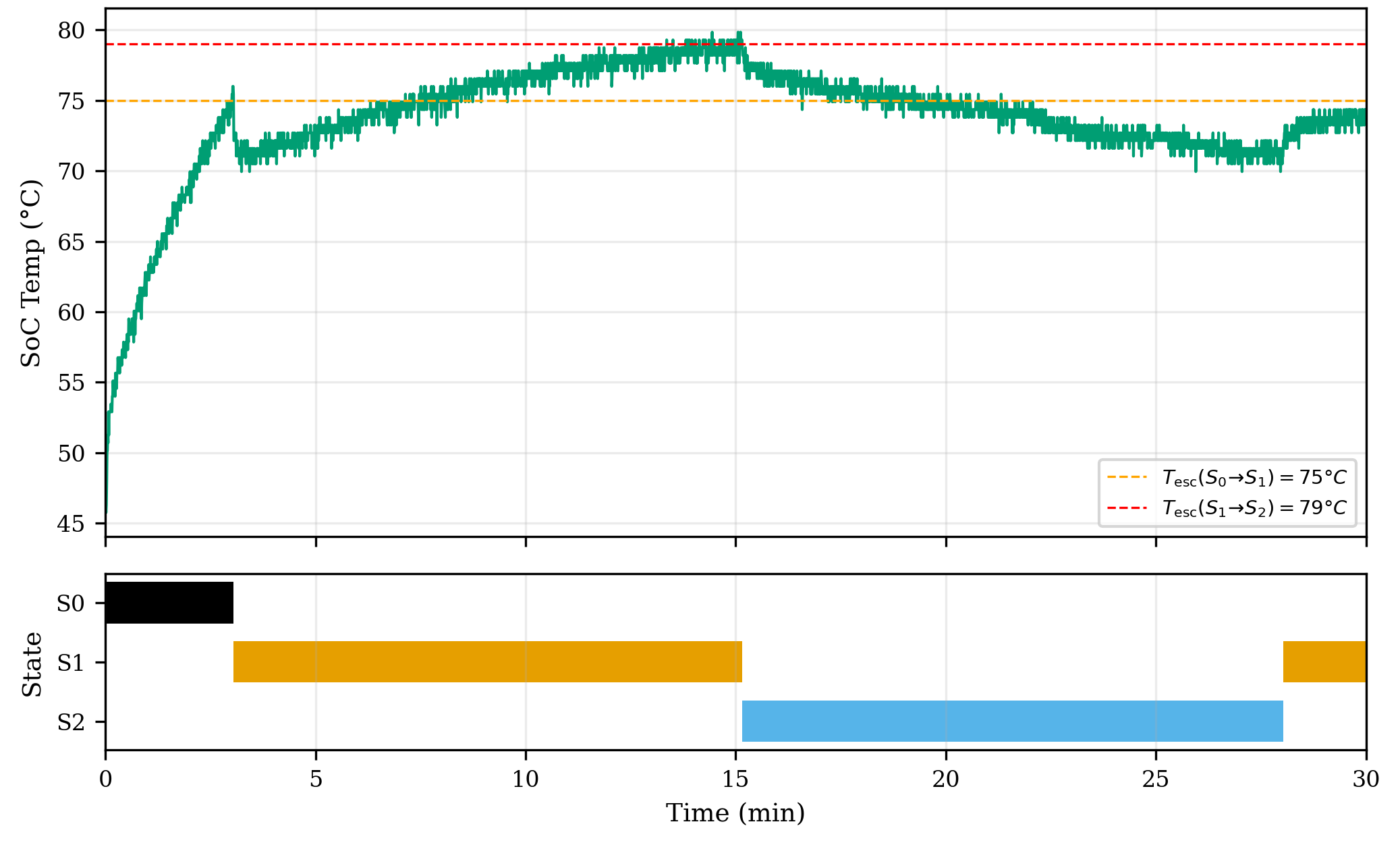}
    \caption{Scheduler Decision Timeline}\label{fig:decision_timeline}
    \end{figure}

Proactive outperformed the reactive baseline on throughput by \textbf{+6.8\%} (Cohen's $d = 8.73$) while using \textbf{1.9\% less energy per frame}, both at zero throttle, versus 1823 events for uncontrolled S0 (Tables~\ref{tab:main_results} and \ref{tab:pairwise_comparison}). Large effect sizes arise primarily from the extremely low run-to-run variance rather than large absolute throughput differences.
The mechanism is time-at-state: Proactive held the efficient S1 state \textbf{54.2\%} of the run while Reactive spent \textbf{0\%} there, locked between S0 and S2 (Fig.~\ref{fig:time_at_state}). The representative Proactive run (Fig.~\ref{fig:decision_timeline}) shows the S0$\to$S1$\to$S2$\to$S1 trajectory: the guards hold each transition, and the scheduler recovers to S1 late in the run as the SoC cools.
All escalations were triggered by the absolute temperature rule. The $\dot{T}$ trigger did not fire at this ambient.
\begin{table}[ht] 
    \centering
    \caption{Static thermal validation}
    \label{tab:thermal_val}
    \begin{tabularx}{\columnwidth}{c c >{\centering\arraybackslash}X c}
        \toprule
        State & T\_plateau ($^\circ$C) & Throttle events & FPS mean \\
        \midrule
        S0 & $84.8 \pm 0.0$ & $1823 \pm 81$ & $12.307 \pm 0.053$ \\
        S1 & $80.3 \pm 0.7$ & $0$ & $12.133 \pm 0.039$ \\
        S2 & $73.4 \pm 0.3$ & $0$ & $10.757 \pm 0.022$ \\
        \bottomrule
    \end{tabularx}
\end{table}
\begin{table*}[htbp] 
    \centering
    \caption{Performance and thermal comparison across configurations. $\pm$ denotes population standard deviation over $n{=}3$ replications.}
    \label{tab:main_results}
    \begin{tabular}{l l c c c c c c}
        \toprule
        \textbf{Group} & \textbf{Method} & \textbf{FPS} & \textbf{Power (W)} & \textbf{J/frame} & \textbf{Throttle} & \textbf{T\_pl} & \textbf{Margin} \\ 
        \midrule
        \textbf{Static} & Static-S0  & $12.307 \pm 0.053$ & $7.215 \pm 0.017$ & $0.586 \pm 0.003$ & 1823 & 84.8 & $-2.8$ \\ 
        & Static-S1  & $12.133 \pm 0.039$ & $6.363 \pm 0.046$ & $0.524 \pm 0.005$ & 0 & 80.3 & $+1.7$ \\ 
        & Static-S2  & $10.757 \pm 0.022$ & $5.637 \pm 0.009$ & $0.524 \pm 0.002$ & 0 & 73.4 & $+8.6$ \\ 
        \midrule
        \textbf{Dynamic} & Reactive-Threshold (threshold-only) & $11.066 \pm 0.003$ & $5.986 \pm 0.009$ & $0.541 \pm 0.001$ & 0 & 73.9 & $+8.1$ \\ 
        & \textbf{Proactive } & \textbf{11.820 $\pm$ 0.070} & \textbf{6.275 $\pm$ 0.050} & \textbf{0.531 $\pm$ 0.004} & \textbf{0} & \textbf{73.7} & \textbf{+8.3} \\ 
        & Proactive-No-Dwell  & $11.929 \pm 0.426$ & $6.392 \pm 0.167$ & $0.536 \pm 0.005$ & 0 & 77.0 & $+5.0$ \\ 
        \midrule
        \textbf{Reference} & Active-cooling reference & $14.530 \pm 0.004$ & $8.183 \pm 0.028$ & $0.563 \pm 0.002$ & 0 & 63.4 & $+18.6$ \\ 
        \bottomrule
    \end{tabular}
\end{table*}
\begin{table}[htbp]
    \centering
    \caption{Pairwise comparison of proactive scheduling against baselines}
    \label{tab:pairwise_comparison}
    \resizebox{\columnwidth}{!}{%
    \begin{tabular}{l c c c c c}
        \toprule
        \textbf{Comparison} & \textbf{$\Delta$FPS} & \textbf{$\Delta$FPS\%} & \textbf{$\Delta$J/fr} & \textbf{$\Delta$J/fr\%} & \textbf{d (FPS)} \\ 
        \midrule
        Proactive vs Reactive-Threshold & $+0.754$ & $+6.8\%$ & $-0.0101$ & $-1.9\%$ & $+8.73$ \\ 
        Proactive vs Static-S1 & $-0.313$ & $-2.6\%$ & $+0.0064$ & $+1.2\%$ & $-3.20$ \\ 
        Proactive vs Static-S2 & $+1.063$ & $+9.9\%$ & $+0.0068$ & $+1.3\%$ & $+9.46$ \\ 
        Proactive vs Static-S0 & $-0.487$ & $-4.0\%$ & $-0.0554$ & $-9.4\%$ & $-23.32$ \\ 
        Proactive vs Proactive-No-Dwell & $-0.109$ & $-0.9\%$ & $-0.0052$ & $-1.0\%$ & $-0.21$ \\ 
        Proactive vs Active-cooling ref & $-2.710$ & $-18.6\%$ & $-0.0323$ & $-5.7\%$ & $-32.59$ \\ 
        \bottomrule
    \end{tabular}%
    }
\end{table}

\subsection{Dwell-Guard Ablation}
Removing the dwell guard left mean throughput essentially unchanged ($-0.9\%$, $d = -0.21$, negligible) but \textbf{destroyed run-to-run consistency}. FPS standard deviation rose from 0.070 to 0.426 \textbf{(6$\times$)} and power SD from 0.050 to 0.167 \textbf{(3.3$\times$)} (std columns of Table~\ref{tab:main_results}; per-run detail in Table~\ref{tab:nodwell}). 
The three runs took divergent trajectories. One was locked in S1 and two escalated to S2 at very different times (602 s vs 1520 s), which is visible as the wide No-Dwell spread in Fig.~\ref{fig:pareto_fps}.
The dwell guard therefore contributes thermal regulation and reproducibility, not throughput. 
\begin{table}[htbp]
    \centering
    \caption{No-dwell ablation per-run results (temperatures in $^\circ$C)}
    \label{tab:nodwell}
    \resizebox{\columnwidth}{!}{%
    \begin{tabular}{c c c c c l}
        \toprule
        \textbf{Rep} & \textbf{Amb.} & \textbf{FPS} & \textbf{T\_pl} & \textbf{T\_max} & \textbf{Transitions} \\
        \midrule
        1 & 24.9 & 12.318 & 77.3 & 77.8 & S0$\to$S1 @168 s; locked S1 \\
        2 & 23.3 & 11.336 & 77.3 & 79.3 & S1 @140 s; S2 @602 s \\
        3 & 24.6 & 12.133 & 76.5 & 79.2 & S1 @189 s; S2 @1520 s \\
        \bottomrule
    \end{tabular}}
\end{table}

\subsection{Pareto Efficiency}
\begin{figure}[htbp]
    \centering
    \includegraphics[width=\linewidth]{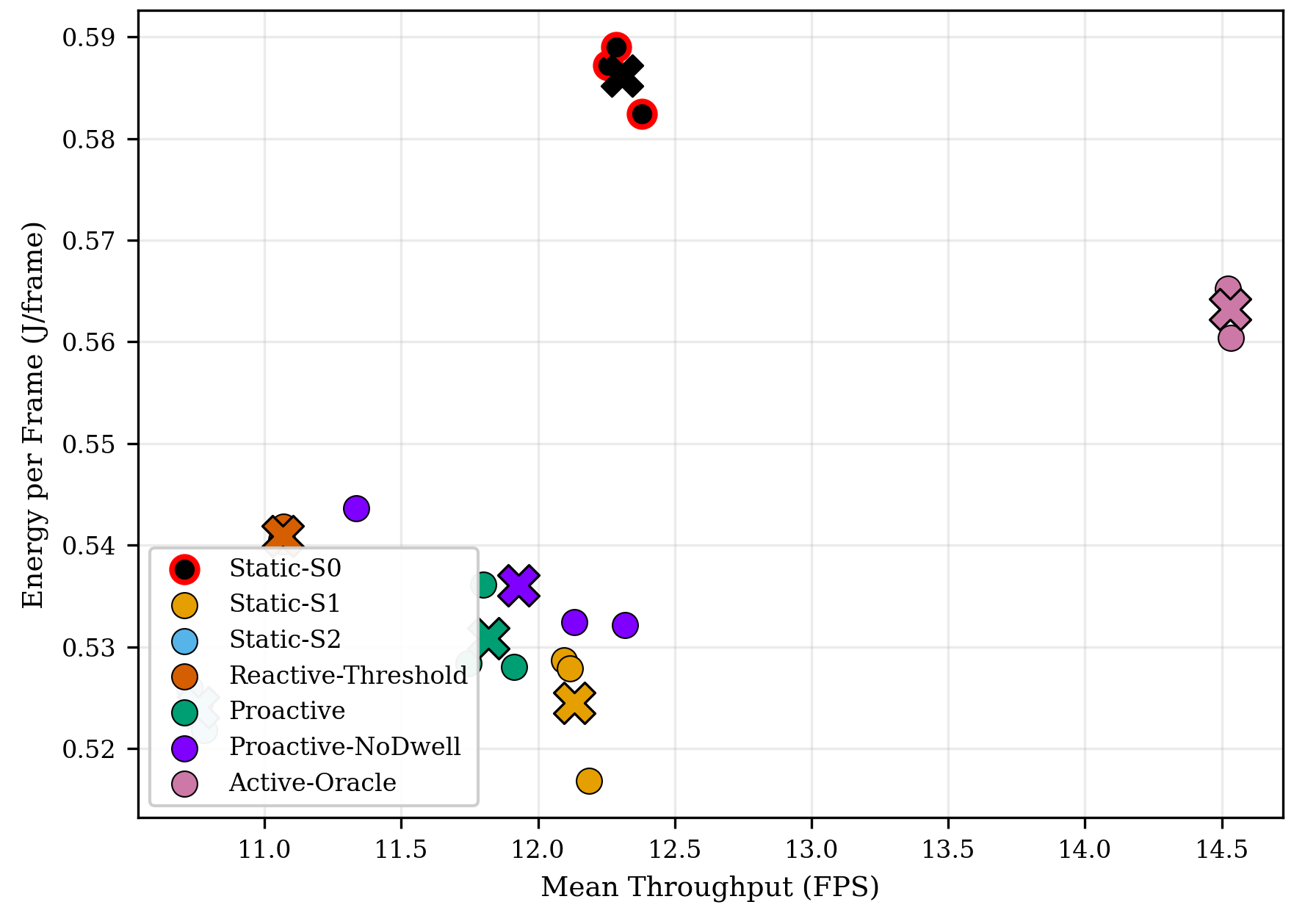}
    \caption{Pareto: Throughput vs Energy (X = mean; red ring = throttle)}\label{fig:pareto_fps}
    \end{figure}

On the throughput-energy plane (Fig.~\ref{fig:pareto_fps}), Proactive sits on the efficient frontier among passive dynamic policies, below Reactive on J/frame and right of Static-S2 on FPS. The active-cooling reference is the fastest but records the highest J/frame among the non-throttling configurations. Only Static S0 throttled (1823 events); all other configurations recorded zero.

\subsection{Sustainability Metrics}
Among dynamic policies, Proactive achieved the best energy per correct detection (Table~\ref{tab:comprehensive_comparison}). The passive Proactive scheduler used less energy per frame than the active cooling reference delivering 81.4\% of its throughput without a fan. It also held an $\qty{8.3}{\celsius}$ safety margin with $\le\qty{1.0}{\celsius}$ overshoot above its escalation target.
\begin{table*}[t]
    \centering
    \caption{Energy efficiency, thermal behavior, and performance across scheduling strategies. J/correct $=$ J/frame $\div$ mAP50 (0.538); mAP is frequency-invariant, so one measurement applies to all conditions. Margin $=$ $\qty{82}{\celsius}$ reference $-$ T\_plateau ($\qty{82}{\celsius}$ is a conservative bound below the $\qty{85}{\celsius}$ throttle limit), in $^\circ$C. Eff\% $=$ FPS relative to the active-cooling reference.}
    \label{tab:comprehensive_comparison}
    \begin{tabular}{lcccccc}
        \toprule
        \textbf{Method} &
        \textbf{J/frame} &
        \textbf{J/correct} &
        \textbf{Throttle exp\%} &
        \textbf{Margin} &
        \textbf{Overshoot} &
        \textbf{Eff\% (vs ref)} \\
        \midrule
        Static-S0 & 0.586 & 1.090 & 50.9\% & $-2.8$ & N/A & 84.7\% \\
        Static-S1 & 0.524 & 0.975 & 0.0\% & $+1.7$ & N/A & 83.5\% \\
        Static-S2 & 0.524 & 0.974 & 0.0\% & $+8.6$ & N/A & 74.0\% \\
        Reactive-Threshold & 0.541 & 1.005 & 0.0\% & $+8.1$ & $-3.0$ & 76.2\% \\
        \textbf{Proactive} & \textbf{0.531} & \textbf{0.987} & \textbf{0.0\%} & \textbf{$+8.3$} & \textbf{$+1.0$} & \textbf{81.4\%} \\
        Proactive-No-Dwell & 0.536 & 0.996 & 0.0\% & $+5.0$ & $+0.5$ & 82.1\% \\
        Active-cooling ref & 0.563 & 1.047 & 0.0\% & $+18.6$ & N/A & 100.0\% \\
        \bottomrule
    \end{tabular}
\end{table*}
\section{Discussion}
\label{sec6}
\subsection{Mechanism}
The observed throughput advantage over the Reactive baseline derives from the \emph{dwell} and $N\_confirm$ safeguards in combination with identical absolute thresholds. These prevent the rapid S1 $\to$ S2 cascade that Reactive executes within its first dwell window. The $\dot{T}$ trigger did not fire in our 30-min nominal ambient runs. The $\dot{T}$-aware decision logic remains in the policy as a safeguard for steeper heating scenarios. Therefore, the core contribution of this work is not the derivative trigger alone, but rather the integrated, empirically calibrated thermal control methodology, where the derivative trigger serves as a necessary safeguard for volatile thermal environments.
 
\subsection{INT8 software-stack finding}
Ancillary profiling of INT8 on this stack (which supports NEON DotProd but lacks the newer I8MM) yielded -43\% FPS and +69.4\% J/frame. Graph analysis of the compiled intermediate representation (IR) explains why: the INT8 model exposes only 64 INT8-typed ports against 842 FP32 ports, $2.22\times$ more FP32 ports than the unquantized FP32 model (379). NNCF inserts Convert (dequantize/requantize) operations at every quantized-layer boundary to bridge INT8 convolutions and FP32 activations, so the DotProd path is largely bypassed. The INT8 model was produced by NNCF 3.0.0 post-training quantization calibrated on the full 481-image validation split (\texttt{fraction=1.0}), exported via
Ultralytics to OpenVINO 2026.0.0 IR, and SHA256-locked as a frozen artifact. This is a finding about this specific runtime and model combination; alternative runtimes may yield different results and are left to future work. So, \textbf{precision optimization is dependent on the software stack; assumed gains can be negative on SoCs without runtime-specific tuning.}

\subsection{Ambient-locked oracle}
 S1's plateau ($\qty{80.3}{\celsius}$) leaves only a $\qty{1.7}{\celsius}$ margin below the $\qty{82}{\celsius}$ reference, versus Proactive's plateau of $\qty{73.7}{\celsius}$ and $\qty{8.3}{\celsius}$ margin (Table~\ref{tab:comprehensive_comparison}). So, although Static S1 edged Proactive on FPS and J/frame, this can be credited to a hardware/cooling coincidence at the specific ambient. A slightly different silicon process variation or even a different heatsink area would likely shift the picture given the minimal margin. Static S1's optimality is ambient-bound: its zero-throttle property vanishes at $\qty{31}{\celsius}$ (Table~\ref{tab:boundary}). An adaptive policy is what we want when ambient varies. 
\subsection{Passive vs active cooling}
As seen in Table~\ref{tab:main_results}, the passive scheduler beats active cooling on J/frame. While the actively cooled reference delivers the maximum raw performance, smart passive scheduling optimizes for efficiency. Smart passive scheduling, in our measurements, achieved lower energy per frame than mechanical cooling. This is attributable to the additional power draw by the fan on the same USB-C bus. Active reference also runs S0 at 2400 MHz the entire time, but proactive runs S1 and a substantial amount of time in S2. A 23.3\% power reduction outweighs the 18.6\% throughput reduction on the J/frame metric. 
\subsection{Why Reactive Cannot Reach S1}
Our reactive baseline locks into S2 by physics, not by implementation: Reactive cascades to S2 early in the run, and the S2 plateau ($\qty{73.4}{\celsius}$--$\qty{73.9}{\celsius}$, Tables~\ref{tab:thermal_val} and \ref{tab:main_results}) settles above the recovery threshold. In every Reactive run, temperature under S2 remained above $T_{\text{rec}}(\mathrm{S2}\rightarrow\mathrm{S1}) = \qty{71}{\celsius}$ for the rest of the run, so the recovery condition never fired and the policy stayed locked in S2.
With identical thresholds, only the guarded policy retains access to S1, which isolates the value of the time-domain guard stack over temperature-threshold-only control.
\section{Limitations}                                                                

\label{sec7}
This study tests a single hardware platform (Pi 5). Generalizability to other DVFS-capable passive SoCs remains future work. We have only tested a single workload and a single model family (YOLOv8n). The model size is expected to affect the layer boundary overhead in the INT8 characterization. The boundary probes in Table~\ref{tab:boundary} were only conducted once for exploratory purposes and require rigorous multi-run validation in future work. The linear thermal model breaks down at ambient temperature $\ge$ $\qty{27}{\celsius}$ due to nonlinear leakage current. To the best of our knowledge, there are no directly comparable open-source implementations available for the given hardware. The temperature-threshold-only reactive baseline serves as the guard-free comparison; because the $\dot{T}$ trigger did not fire at nominal ambient, the derivative's individual contribution is not isolated by these experiments. While this study highlights a severe quantization penalty when deploying via OpenVINO on the Cortex-A76, this exposes a critical dependency on software-stack maturity rather than an absolute hardware limit. Runtime dependence is left for future work.
\begin{table}[htbp] 
    \centering
    \caption{Exploratory boundary-probe runs}
    \label{tab:boundary}
    \begin{tabularx}{\columnwidth}{c c c c >{\raggedright\arraybackslash}X}
        \toprule
        \textbf{Ambient} & \textbf{Method} & \textbf{FPS} & \textbf{Peak T} & \textbf{Outcome} \\ 
        \midrule
        23 $^\circ$C & Proactive & 11.820 & 79 $^\circ$C & no throttle \\ 
        27 $^\circ$C & Proactive & 10.92 & $\approx$89 $^\circ$C & throttle takeover \\ 
        31 $^\circ$C & Proactive & 9.55 & $\approx$89 $^\circ$C & kernel forced reduction \newline (throttle\_raw = 0xE0006) \\ 
        31 $^\circ$C & Static-S1 & 11.08 & $\approx$88 $^\circ$C & heavy throttle \\ 
        \bottomrule
    \end{tabularx}
\end{table}

\textbf{Linear model deviation:} predicted S2 plateau at 31 $^\circ$C = 81 $^\circ$C; observed 88--89 $^\circ$C $\rightarrow$ nonlinear leakage at elevated junction temp. 
\section{Conclusion}
\label{sec8}
In this research, we developed and evaluated an empirically calibrated state-aware DVFS thermal control methodology for sustained inference on passively cooled edge SoCs, with every guard parameter derived from on-device sensor-noise and step-response measurements rather than heuristic tuning, validated under a reproducible protocol. The integrated policy outperformed temperature-only reactive thresholding on both throughput and energy efficiency while eliminating all observed thermal throttling, and the dwell guard proved necessary for run-to-run reproducibility at a negligible ($<1\%$) throughput cost. Exploratory probes indicate the passive envelope closes at ambient $\ge \qty{27}{\celsius}$ where nonlinear leakage drives even the lowest DVFS state (S2) above the throttle threshold. Therefore, while correct scheduling can make mechanical cooling unnecessary on this platform, this holds strictly within this mapped ambient operating envelope. Future work includes additional DVFS-capable platforms and workloads, runtime-specific INT8 evaluation, and model-predictive extensions of the control policy.

\bibliographystyle{IEEEtran}
\bibliography{references}

@inproceedings{ahmadiEdgeEngineThermalAwareOptimization2024,
  title = {{{EdgeEngine}}: {{A Thermal-Aware Optimization Framework}} for {{Edge Inference}}},
  shorttitle = {{{EdgeEngine}}},
  booktitle = {Proceedings of the {{Eighth ACM}}/{{IEEE Symposium}} on {{Edge Computing}}},
  author = {Ahmadi, Amirhossein and Abdelhafez, Hazem A. and Pattabiraman, Karthik and Ripeanu, Matei},
  year = 2024,
  month = aug,
  series = {{{SEC}} '23},
  pages = {67--79},
  publisher = {Association for Computing Machinery},
  address = {New York, NY, USA},
  doi = {10.1145/3583740.3626616},
  urldate = {2026-06-11},
  isbn = {979-8-4007-0123-8}
}

@misc{ahnPerformanceCharacterizationUsing2023,
  title = {Performance {{Characterization}} of Using {{Quantization}} for {{DNN Inference}} on {{Edge Devices}}: {{Extended Version}}},
  shorttitle = {Performance {{Characterization}} of Using {{Quantization}} for {{DNN Inference}} on {{Edge Devices}}},
  author = {Ahn, Hyunho and Chen, Tian and Alnaasan, Nawras and Shafi, Aamir and Abduljabbar, Mustafa and Subramoni, Hari and Panda, Dhabaleswar K.},
  year = 2023,
  month = mar,
  number = {arXiv:2303.05016},
  eprint = {2303.05016},
  primaryclass = {cs.PF},
  publisher = {arXiv},
  doi = {10.48550/arXiv.2303.05016},
  urldate = {2026-06-11},
  archiveprefix = {arXiv}
}

@misc{aryaRDD2022MultinationalImage2022,
  title = {{{RDD2022}}: {{A}} Multi-National Image Dataset for Automatic {{Road Damage Detection}}},
  shorttitle = {{{RDD2022}}},
  author = {Arya, Deeksha and Maeda, Hiroya and Ghosh, Sanjay Kumar and Toshniwal, Durga and Sekimoto, Yoshihide},
  year = 2022,
  month = sep,
  number = {arXiv:2209.08538},
  eprint = {2209.08538},
  primaryclass = {cs.CV},
  publisher = {arXiv},
  doi = {10.48550/arXiv.2209.08538},
  urldate = {2026-06-11},
  archiveprefix = {arXiv}
}

@misc{ashfaqAcceleratingDeepLearning2022,
  title = {Accelerating {{Deep Learning Model Inference}} on {{Arm CPUs}} with {{Ultra-Low Bit Quantization}} and {{Runtime}}},
  author = {Ashfaq, Saad and AskariHemmat, MohammadHossein and Sah, Sudhakar and Saboori, Ehsan and Mastropietro, Olivier and Hoffman, Alexander},
  year = 2022,
  month = jul,
  number = {arXiv:2207.08820},
  eprint = {2207.08820},
  primaryclass = {cs.LG},
  publisher = {arXiv},
  doi = {10.48550/arXiv.2207.08820},
  urldate = {2026-06-11},
  archiveprefix = {arXiv}
}

@manual{aosong_dht11,
  author       = {{Aosong (Guangzhou) Electronics Co., Ltd.}},
  title        = {DHT11 Product Manual: Temperature and Humidity Module},
  organization = {Aosong (Guangzhou) Electronics Co., Ltd.},
  url          = {http://www.aosong.com},
  note         = {Accessed: 2026-06-18}
}

@article{jacobThermoAwareAnytimeInference,
  title = {Thermo-{{Aware Anytime Inference}} on {{Raspberry Pi}} 5: {{Uncertainty-Aware Control}} and {{Full-Stack Reproducibility}}},
  shorttitle = {Thermo-{{Aware Anytime Inference}} on {{Raspberry Pi}} 5},
  author = {Jacob, Manu Nicholas},
  journal = {TechRxiv},
  year = {2025},
  number = {1030},
  publisher = {TechRxiv},
  doi = {10.36227/techrxiv.176186797.71232313/v1},
  urldate = {2026-06-11}
}

@article{jainConvolutionalNeuralNetworks2021,
  title = {Convolutional Neural Networks for Real-Time Object Detection with Raspberry {{Pi}}},
  author = {Jain, Mohit and Shah, Adit},
  year = 2021,
  month = dec,
  journal = {World Journal of Advanced Engineering Technology and Sciences},
  volume = {4},
  pages = {87--105},
  doi = {10.30574/wjaets.2021.4.1.0067}
}

@article{jeonPhoenixThermalAwareOnDevice2026,
  title = {Phoenix: {{Thermal-Aware On-Device Inference}} of {{Multi-Instance DNNs}} for {{Mobile Video Applications}}},
  shorttitle = {Phoenix},
  author = {Jeon, Seunghyeok and Kim, Jiwon and Lee, Jeho and Cha, Hojung},
  year = 2026,
  month = mar,
  journal = {ACM Trans. Embed. Comput. Syst.},
  volume = {25},
  number = {2},
  pages = {31:1--31:23},
  issn = {1539-9087},
  doi = {10.1145/3793860},
  urldate = {2026-06-11}
}

@inproceedings{koseBridgingGapAI2025,
  title = {Bridging the {{Gap Between AI Quantization}} and {{Edge Deployment}}: {{INT4}} and {{INT8}} on the {{Edge}}},
  shorttitle = {Bridging the {{Gap Between AI Quantization}} and {{Edge Deployment}}},
  booktitle = {5th {{Muslims}} in {{ML Workshop}} Co-Located with {{NeurIPS}} 2025},
  author = {K{\"o}se, Mohammad Ibrahim and Ahmed, Qazi Arbab and Jungeblut, Thorsten},
  year = 2025,
  month = nov,
  urldate = {2026-06-11},
  langid = {english}
}

@misc{liEnergyEfficientComputationDVFS2025,
  title = {Energy-{{Efficient Computation}} with {{DVFS}} Using {{Deep Reinforcement Learning}} for {{Multi-Task Systems}} in {{Edge Computing}}},
  author = {Li, Xinyi and Zhou, Ti and Wang, Haoyu and Lin, Man},
  year = 2025,
  month = may,
  number = {arXiv:2409.19434},
  eprint = {2409.19434},
  primaryclass = {cs.OS},
  publisher = {arXiv},
  doi = {10.48550/arXiv.2409.19434},
  urldate = {2026-06-11},
  archiveprefix = {arXiv}
}

@article{mahmudovQuantEdgeHybridQuantization2025,
  title = {{{QuantEdge}}: {{A Hybrid Quantization Approach}} for {{Optimized AI Deployment Across Edge Devices}}},
  shorttitle = {{{QuantEdge}}},
  author = {Mahmudov, Rasim and Kim, Deok-Hwan},
  year = 2025,
  journal = {IEEE Access},
  volume = {13},
  pages = {161605--161618},
  issn = {2169-3536},
  doi = {10.1109/ACCESS.2025.3609798},
  urldate = {2026-06-11}
}

@article{ngoEdgeIntelligenceReview2025,
  title = {Edge {{Intelligence}}: {{A Review}} of {{Deep Neural Network Inference}} in {{Resource-Limited Environments}}},
  shorttitle = {Edge {{Intelligence}}},
  author = {Ngo, Dat and Park, Hyun-Cheol and Kang, Bongsoon},
  year = 2025,
  month = jan,
  journal = {Electronics},
  volume = {14},
  number = {12},
  pages = {2495},
  publisher = {Multidisciplinary Digital Publishing Institute},
  issn = {2079-9292},
  doi = {10.3390/electronics14122495},
  urldate = {2026-06-11},
  copyright = {http://creativecommons.org/licenses/by/3.0/},
  langid = {english}
}

@inproceedings{nishaRunTimePreventionThermal2024,
  title = {Run-{{Time Prevention}} of {{Thermal Throttling}} on the {{Edge}} Using {{Reinforcement-Learning Based Predictive Thermal Aware Power}} and {{Performance Management}}},
  booktitle = {2024 22nd {{IEEE Interregional NEWCAS Conference}} ({{NEWCAS}})},
  author = {Nisha, Parveen and Vinay, Ratnala and Laad, Kartik and Sasmal, Pradip and Haraki, Toshihisa and Juyal, Chirag and Gabir Elbakri, Mohamed Amir and Acharyya, Amit},
  year = 2024,
  month = jun,
  pages = {273--277},
  issn = {2474-9672},
  doi = {10.1109/NewCAS58973.2024.10666109},
  urldate = {2026-06-11}
}

@article{nizeniecYOLOObjectDetectors2026,
  title = {{{YOLO Object Detectors}} for {{Robotics}} -- a {{Comparative Study}}},
  author = {Ni{\.z}eniec, Patryk and Iwanowski, Marcin and Gahbler, Marcin},
  year = 2026,
  month = mar,
  journal = {PAR},
  volume = {30},
  number = {1},
  eprint = {2603.27029},
  primaryclass = {cs.CV},
  pages = {117--126},
  issn = {14279126},
  doi = {10.14313/PAR_259/117},
  urldate = {2026-06-11},
  archiveprefix = {arXiv}
}

@inproceedings{pallipadiOndemandGovernor2006,
  author    = {Pallipadi, Venkatesh and Starikovskiy, Alexey},
  title     = {The Ondemand Governor: Past, Present, and Future},
  booktitle = {Proceedings of the Linux Symposium},
  volume    = {2},
  pages     = {215--230},
  year      = {2006},
  month     = {July},
  address   = {Ottawa, Ontario, Canada},
  url       = {https://www.kernel.org/doc/ols/2006/ols2006v2-pages-223-238.pdf}
}

@misc{ultralytics_yolov8,
  author       = {Jocher, Glenn and Chaurasia, Ayush and Qiu, Jing},
  title        = {Ultralytics {YOLOv8}},
  year         = {2023},
  publisher    = {GitHub},
  journal      = {GitHub repository},
  howpublished = {\url{https://github.com/ultralytics/ultralytics}},
  note         = {Version 8.4.7}
}

@article{shuvoEfficientAccelerationDeep2022,
  title = {Efficient {{Acceleration}} of {{Deep Learning Inference}} on {{Resource-Constrained Edge Devices}}: {{A Review}}},
  shorttitle = {Efficient {{Acceleration}} of {{Deep Learning Inference}} on {{Resource-Constrained Edge Devices}}},
  author = {Shuvo, Md. Maruf Hossain and Islam, Syed and Cheng, Jianlin and Morshed, Bashir},
  year = 2022,
  month = dec,
  journal = {Proceedings of the IEEE},
  volume = {111},
  pages = {1--50},
  doi = {10.1109/JPROC.2022.3226481}
}

@article{yatskivPREDICTIVETHERMALMANAGEMENT2026,
  title = {{{PREDICTIVE THERMAL MANAGEMENT IN EMBEDDED ELECTRONICS USING DEEP REINFORCEMENT LEARNING}}},
  author = {Yatskiv, Oleh and Koman, Bohdan},
  year = 2026,
  month = apr,
  journal = {Electronics and Information Technologies},
  number = {33},
  pages = {145--164},
  issn = {2224-0888},
  doi = {10.30970/eli.33.11},
  urldate = {2026-06-11},
  copyright = {Copyright (c) 2026 Oleh Yatskiv, Bohdan Koman},
  langid = {english}
}

@article{zhangE4EnergyEfficientDNN2025,
  title = {E4: {{Energy-Efficient DNN Inference}} for {{Edge Video Analytics}} via {{Early Exiting}} and {{DVFS}}},
  shorttitle = {E4},
  author = {Zhang, Ziyang and Zhao, Yang and Chang, Ming-Ching and Lin, Changyao and Liu, Jie},
  year = 2025,
  month = apr,
  journal = {Proceedings of the AAAI Conference on Artificial Intelligence},
  volume = {39},
  number = {1},
  pages = {1165--1173},
  issn = {2374-3468},
  doi = {10.1609/aaai.v39i1.32104},
  urldate = {2026-06-11},
  copyright = {Copyright (c) 2025 Association for the Advancement of Artificial Intelligence},
  langid = {english}
}

@misc{zhangSparseDVFSSparseAwareDVFS2026,
  title = {{{SparseDVFS}}: {{Sparse-Aware DVFS}} for {{Energy-Efficient Edge Inference}}},
  shorttitle = {{{SparseDVFS}}},
  author = {Zhang, Ziyang and Wu, Zheshun and Liu, Jie and Mottola, Luca},
  year = 2026,
  month = mar,
  number = {arXiv:2603.21908},
  eprint = {2603.21908},
  primaryclass = {cs.LG},
  publisher = {arXiv},
  doi = {10.48550/arXiv.2603.21908},
  urldate = {2026-06-11},
  archiveprefix = {arXiv}
}

@article{zhangThermalAwareOnDeviceInference2022,
  title = {Thermal-{{Aware}} on-{{Device Inference Using Single-Layer Parallelization}} with {{Heterogeneous Processors}}},
  author = {Zhang, Jinghui and Wang, Yuchen and Huang, Tianyu and Dong, Fang and Zhao, Wei and Shen, Dian},
  year = 2022,
  month = jul,
  journal = {Tsinghua Science and Technology},
  volume = {28},
  pages = {82--92},
  doi = {10.26599/TST.2021.9010075}
}

\end{document}